\documentclass[aps,prl,twocolumn,groupedaddress]{revtex4-1}
\usepackage{hyperref}
\usepackage{graphicx}
\usepackage{dcolumn}

\def\aaa{Astron.\ Astroph.\ }

\def\apj{Astrophys.\ J. }
\def\apjl{Astrophys.\ J.\ Lett.\  }

\def\mnras{Month.\ Not.\ Roy.\ Astron.\ Soc.\ }

\def\prd{Phys.\ Rev.\ D }

\def\prl{Phys.\ Rev.\ Lett.\ }

\def\jms{J. Mol.\ Spectrosc.\ }

\usepackage{verbatim}

\newcolumntype{.}{D{.}{.}{3.7}}

\begin{document}


\title{Robust Constraint on a Drifting Proton-to-Electron Mass Ratio at $z=0.89$ \\
from Methanol Observation at Three Radio Telescopes}



\author{J. Bagdonaite, M. Dapr\`{a}, P. Jansen, H. L. Bethlem, W. Ubachs}
\affiliation{Department of Physics and Astronomy, VU University Amsterdam, De Boelelaan 1081, 1081 HV Amsterdam, The Netherlands}
\author{S. Muller}
\affiliation{Onsala Space Observatory, Chalmers University of Technology, SE 439-92, Onsala, Sweden}
\author{C. Henkel}
\affiliation{Max-Planck-Institut f\"{u}r Radioastronomie, Auf dem H\"{u}gel 69, 53121 Bonn, Germany}
\affiliation{Astronony Department, King Abdulaziz University, Post Office Box 80203, Jeddah 21589, Saudi Arabia}
\author{K. M. Menten}
\affiliation{Max-Planck-Institut f\"{u}r Radioastronomie, Auf dem H\"{u}gel 69, 53121 Bonn, Germany}


\begin{abstract}
A limit on a possible cosmological variation of the proton-to-electron mass ratio $\mu$ is derived from methanol (CH$_3$OH) absorption lines in the benchmark PKS1830$-$211 lensing galaxy at redshift $z = 0.89$ observed with the Effelsberg 100-m radio telescope, the Institute de Radio Astronomie Millim\'{e}trique 30-m telescope, and the Atacama Large Millimeter/submillimeter Array. Ten different absorption lines of CH$_3$OH covering a wide range of sensitivity coefficients $K_{\mu}$ are used to derive a purely statistical 1-$\sigma$ constraint of $\Delta\mu/\mu = (1.5 \pm 1.5) \times 10^{-7}$ for a lookback time of 7.5 billion years. Systematic effects of chemical segregation, excitation temperature, frequency dependence and time variability of the background source are quantified. A multi-dimensional linear regression analysis leads to a robust constraint of $\Delta\mu/\mu = (-1.0 \pm 0.8_{\rm stat} \pm 1.0_{\rm sys}) \times 10^{-7}$.
\end{abstract}

\maketitle

The fact that the strengths of the fundamental forces of nature are not fixed by the Standard Model of physics leaves space for the hypothesis of coupling constants varying over time and space. Such a constant is the dimensionless proton-to-electron mass ratio $\mu=m_{\rm p}/m_{\rm e}$, representing the ratio of the strong force to the electroweak scale~\cite{Flambaum2004}. Effects of a time-varying $\mu$ may be probed through the measurement of molecular line spectra in various cosmological epochs. The spectrum of molecular hydrogen, H$_2$, the most abundant molecule in the Universe, can be investigated with large optical telescopes~\cite{King2008,Malec2010,Weerdenburg2011,Wendt2012}.
The sensitivity coefficient $K_{\mu}$, defined via
\begin{equation}
   \frac{\Delta\nu}{\nu}= K_{\mu} \frac{\Delta\mu}{\mu}
\label{eq1}
\end{equation}
for the H$_2$ spectral lines is only of the order of $0.05$~\cite{Ubachs2007}.
For this reason the $\mu$ constraint resulting from H$_2$ data is not better than $\Delta\mu/\mu < 10^{-5}$. 

Transitions of some other molecules in the radio part of the electromagnetic spectrum exhibit larger sensitivities to a varying $\mu$. Inversion transitions of ammonia (NH$_3$), with $K_{\mu}=-4.46$, have been applied to produce 1-$\sigma$ constraints at the level of $\Delta\mu/\mu=(1.0 \pm 4.7) \times 10^{-7}$ in the object PKS1830$-$211 at $z=0.89$~\cite{Henkel2009} and $(-3.5 \pm 1.2) \times 10^{-7}$ in the object B0218$+$357 at $z=0.68$~\cite{Kanekar2011}. In the $\mu$-variation analysis the strongly shifting NH$_3$ lines must be compared with non-shifting anchor lines belonging to different species, such as HCO$^+$ and HC$_3$N. This may give rise to systematic effects on the result for $\Delta\mu/\mu$ due to chemical segregation, \emph{i.e.} to a non-homogeneous spatial distribution of the various molecular species along the line-of-sight.

It was recently pointed out that the interplay between the internal and overall rotation in the methanol molecule (CH$_3$OH) results in specific transitions having an enhanced sensitivity for a possible drift in $\mu$~\cite{Jansen2011a,Levshakov2011}. Some of these transitions involve low lying rotational energy levels populated at the low temperatures characterizing the bulk of the interstellar molecular gas. The spread in $K_{\mu}$ coefficients for methanol lines provides the unique opportunity of deriving a tight constraint on $\mu$ from a single molecular species, therewith avoiding chemical segregation issues. 
Methanol has recently been observed in a gravitationally lensing galaxy (at $z=0.89$, corresponding to a lookback time of 7.5 billion years \cite{hubble}) toward the south-western  (SW) image of the background blazar PKS1830$-$211~\cite{Muller2011,Ellingsen2012}. In a preliminary investigation based on a small sample of four methanol transitions a 1-$\sigma$ limit of $\Delta\mu/\mu$ at 1$\times 10^{-7}$ was derived~\cite{Bagdonaite2013}, with an indication of a spatial differentiation of the $E$- and $A$-type symmetry species. Here, we present an extended study of $\mu$ variation based on 17 measurements of ten different absorption lines of CH$_3$OH allowing for a quantitative analysis of previously unaddressed underlying systematic effects. This analysis leads to a similar constraint as the one found in \cite{Bagdonaite2013}, but the robustness of this constraint is greatly improved. 

\begin{figure*}
\includegraphics[scale=.25, trim= 0cm 2.5cm 0cm 1cm]{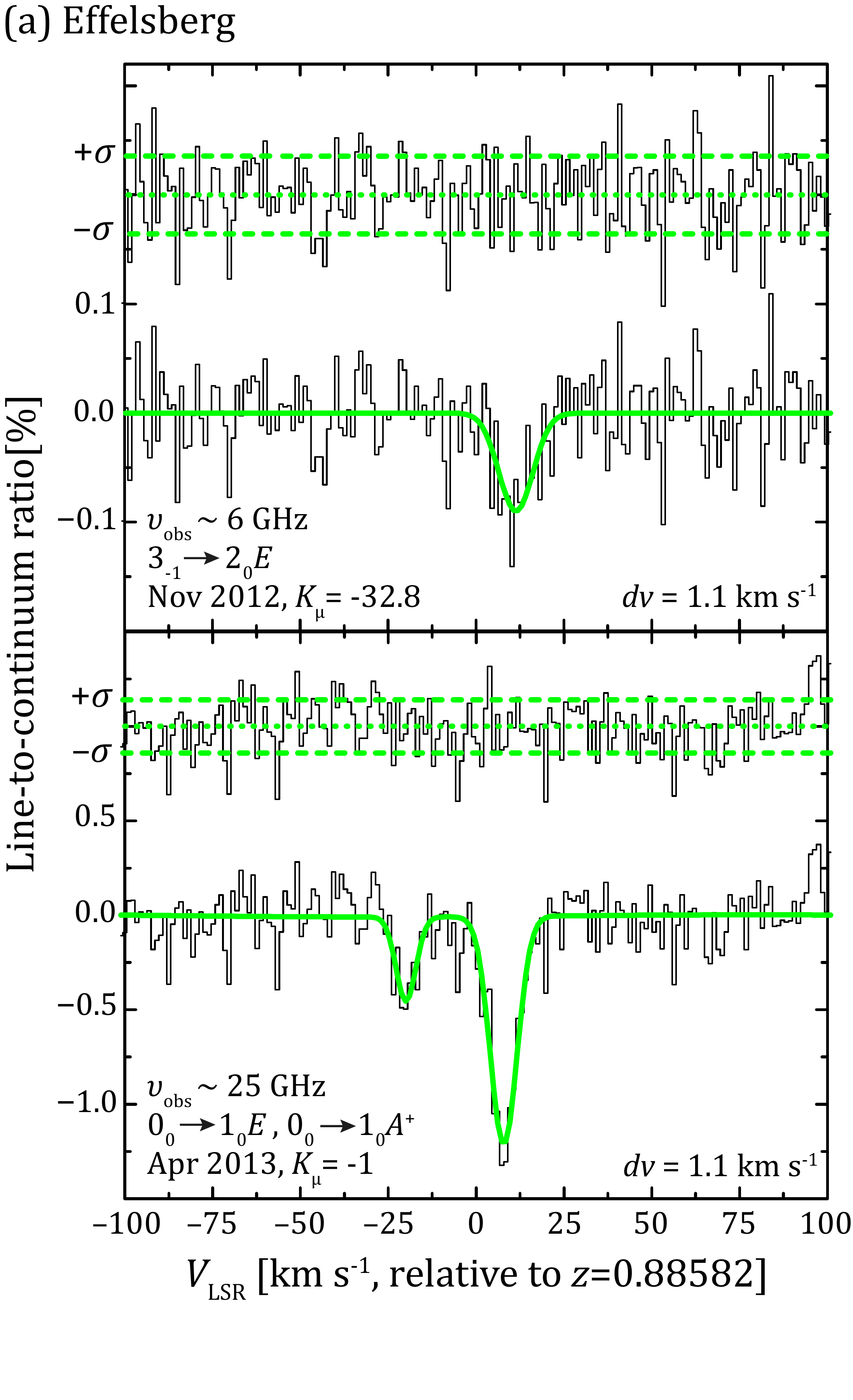}
\includegraphics[scale=.25, trim= 0cm 2.5cm 0cm 1cm ]{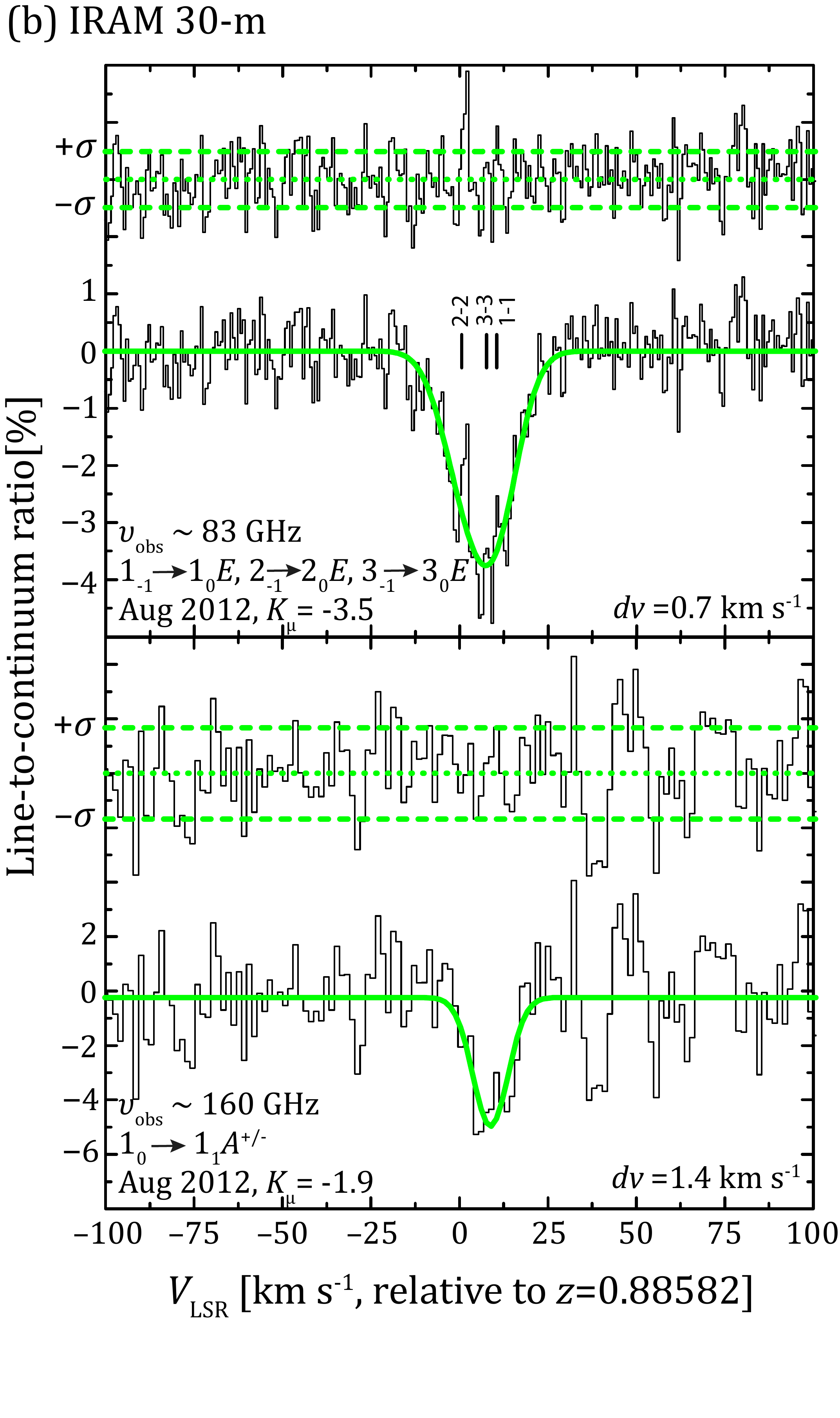}
\includegraphics[scale=.25, trim= 0cm -3cm 0cm 0cm ]{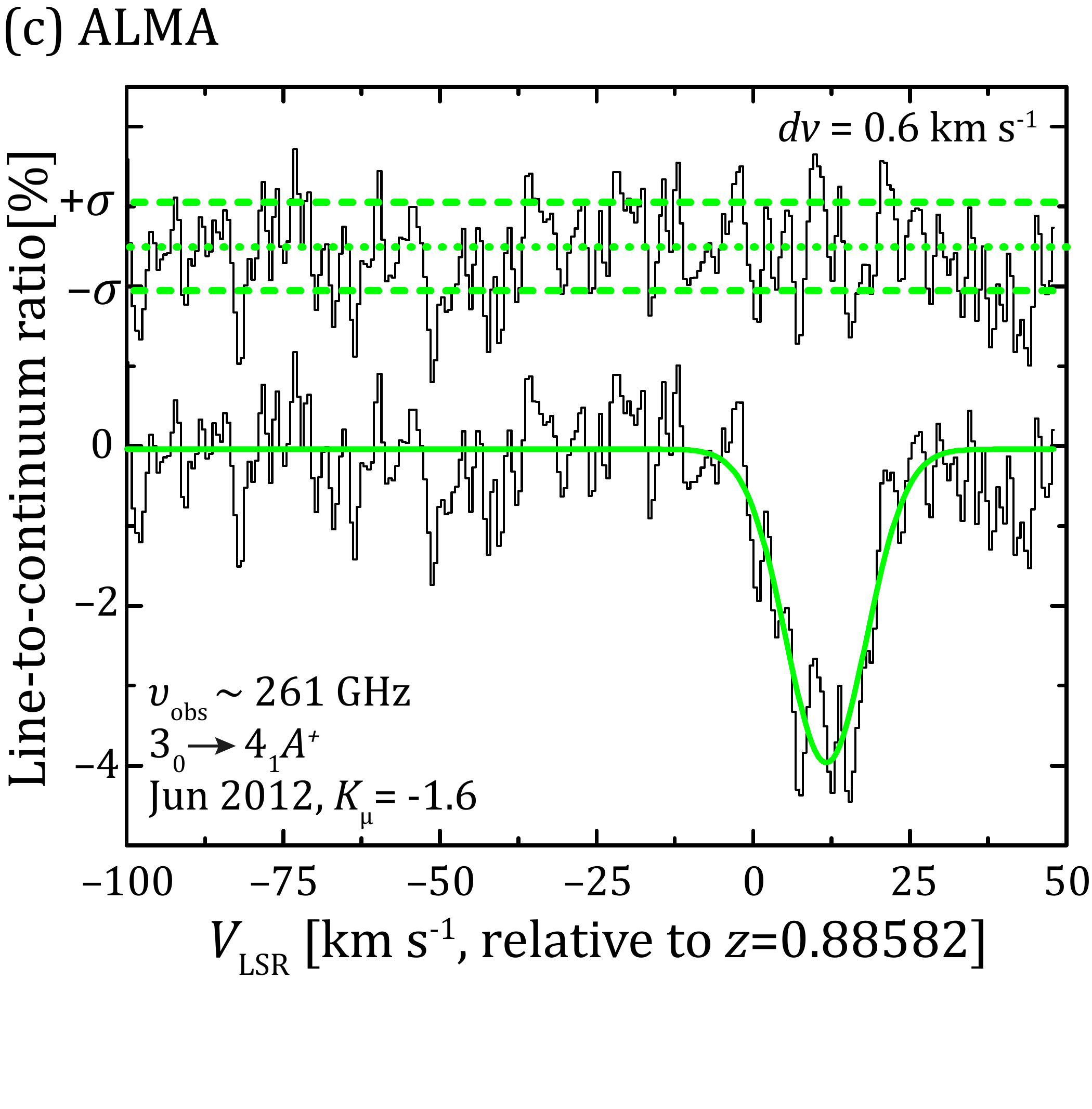}
\caption{\small{Some of the observed methanol transitions toward PKS1830$-$211 with their approximate observed frequencies $\nu_{\rm obs}$, observation epoch, sensitivity coefficients $K_{\mu}$, and channel spacings $dv$ indicated in each panel for recordings at (a) Effelsberg radio telescope, (b) IRAM 30-m telescope, (c) ALMA array. The intensity is normalized to the total continuum flux, except for (c) where the SW image could be observed exclusively. Gaussian fits to the absorption profiles are displayed in light green with residuals shown at the top of each panel.
}}
\label{Fig1}
\end{figure*}

Methanol absorption spectra were recorded using three different radio telescopes covering a range of 6--261 GHz. 
The CH$_3$OH lines in the low-frequency range (detected frequencies $< 35$ GHz), previously observed in the time slot Dec. 2011--Apr. 2012 \cite{Bagdonaite2013}, were reobserved with the 100-m single-dish Effelsberg radio telescope of the Max-Planck-Institute f\"{u}r Radioastronomie 
(see Table~\ref{tbl-1}). Spectra of the new observations in fall 2012 and spring 2013 are shown in Fig.~\ref{Fig1}(a). The absorption profiles of individual transitions were fitted to single Gaussians, and the line positions and widths were determined on a Local Standard of Rest (LSR) velocity scale, which was centered at $z = 0.88582$.  

The Institute de Radio Astronomie Millim\'{e}trique (IRAM) 30-m single dish telescope at Pico Veleta (Spain) was used to detect methanol lines in the range 80$-$165 GHz, using the EMIR receiver in observations during 22$-$27 August 2012. Fig.~\ref{Fig1}(b) shows the overlapping components of the  $1_{-1}-1_0E$,  $2_{-1}-2_0E$ and  $3_{-1}-3_0E$ transitions at a detection frequency of 83.3 GHz. A fit was performed assuming that relative intensities between the three components are proportional to their expected optical depths. The RADEX radiative transfer model \citep{vandertak2007} was used to estimate the relative intensities. Also the relative positions of the components were fixed based on the laboratory measurements, and the widths were assumed to be the same. The uncertainty of the combined line position is estimated by varying the relative strengths of the three lines by 10-20$\%$ with regard to the RADEX prediction. Also shown in Fig.~\ref{Fig1} is a recording of the $1_0-1_1A^{+/-}$ line at a detection frequency of 160 GHz. In addition the weak $3_0 - 2_1A^+$ line at $\nu_{\rm obs} \sim 83.0$ GHz was observed with the IRAM 30-m.

An observation of the $3_0-4_1A^+$ CH$_3$OH transition at a frequency of 261 GHz was conducted in June 2012 with the Atacama Large Millimeter/submillimeter Array [ALMA, 20 dishes, 12 m in diameter each; see Fig. \ref{Fig1}(c)]. The angular resolution of the ALMA observations ($\sim$0.6 arcsec) made it possible to obtain a spectrum toward the south-western image only.


\begin{table*}
\begin{center}
\caption{A summary of the observed methanol lines in PKS1830$-$211. Listed are the lower and upper energy level quantum numbers, laboratory or rest frequencies $\nu_{\rm Lab}$ and corresponding uncertainties, fractional uncertainties $\Delta\nu/\nu$, uncertainties in terms of Doppler shift, $\Delta v_{\rm D}$, and corresponding frequencies at redshift $z=0.88582$. $K_{\mu}$ refers to the sensitivity coefficients (see Eq. \ref{eq1}). Fitted line positions $V_{\rm LSR}$ are stated with respect to the Local Standard of Rest (centered at $z=0.88582$). Uncertainties in position and FWHM width are obtained from Gaussian fits (1$\sigma$). Optical depths $\tau$ are fitted from the spectra. In the last column the instrument and the observation period is mentioned. Eff. refers to the Effelsberg 100-m, IRAM to the 30-m radio telescope at the Pico Veleta.
\label{tbl-1}}
\begin{tabular}{l l l c . l c c l l}
\tableline\tableline
Transition & \multicolumn{1}{c}{$\nu_{\rm Lab}$}  & \multicolumn{1}{c}{$\Delta\nu/\nu$} & \multicolumn{1}{c}{$\Delta v_{\rm D}$}  & \multicolumn{1}{c}{$\nu_{\rm z=0.88582}$} & \multicolumn{1}{l}{$K_{\mu}$} & \multicolumn{1}{c}{$V_{\rm LSR}$}   & \multicolumn{1}{c}{Width}  & \multicolumn{1}{c}{$\tau$}  &\multicolumn{1}{l}{Observation} \\
 \multicolumn{1}{l}{lower$-$upper}& \multicolumn{1}{c}{(GHz)} & \multicolumn{1}{l}{} & \multicolumn{1}{c}{(km s$^{-1}$)} & \multicolumn{1}{c}{(GHz)} &  & \multicolumn{1}{c}{(km s$^{-1}$)}  &  \multicolumn{1}{c}{(km s$^{-1}$)} &\multicolumn{1}{l}{}  & \multicolumn{1}{l}{} \\
\tableline
 $3_{-1}-2_0E$ & 12.178597(4)$^{a}$ & $3 \times 10^{-7}$  & 0.1 & 6.457985 & -32.8   &$9.1\pm 0.7$ & $18.2\pm 1.5$  &0.0024 & Eff. Feb. 2012$^h$\\
 &             &                     &                     &    &           &$10.7\pm 0.7$& $12.0 \pm 1.2$ & 0.002 & Eff. Nov. 2012 \\
 &             &                     &                     &    &           &$12.6\pm 2.0$& $19.9 \pm 4.5$ &0.002  & Eff. May 2013 \\
 &             &                     &                     &    &           &$7.4\pm 1.3^f$& $17.0\pm 2.9^f$ &0.005$^f$  & ATCA Nov. 2011 \\
 $0_0-1_0A^+$  & 48.3724558(7)$^{b}$ & $2 \times 10^{-8}$  & 0.006& 25.6506219 &-1.0  &$8.3\pm 0.1$ & $12.0\pm 0.2$  &0.045  & Eff. Dec. 2011$^h$\\
    &  &  &  &   &							   & $8.8\pm 0.2$  & $14.8\pm 0.6$  & 0.03 & Eff. Apr. 2012$^h$\\
    &  &  &  &   &							   & $8.7\pm 0.2$  & $8.5\pm 0.6$  & 0.03 & Eff. Mar. 2013\\
    &  &  &  &   &							   &  $7.8\pm 0.3$ & $9.0\pm 0.7$ & 0.03 & Eff. Apr. 2013\\
 $0_0-1_0E$    & 48.376892(10)$^{c}$ & $2 \times 10^{-7}$  & 0.06 &25.652974 &-1.0  &$8.9\pm 0.3$ & $18.5\pm 0.8$  &0.016 & Eff. Dec. 2011$^h$\\
   &  &  &  &   &						  	   & $10.4\pm 0.7$  & $12.5\pm 1.6$   & 0.011 &Eff. Apr. 2012$^h$\\
   &  &  &  &	&							   &$7.6\pm 0.6$  &$6.7\pm 1.3$  & 0.013 & Eff. Apr. 2013\\
 $2_{-1}-1_0E$ & 60.531489(10)$^{c}$ & $2 \times 10^{-7}$  & 0.06 &32.098233 &-7.4  &$9.8\pm 0.4$ & $20.0\pm 1.1$  &0.028 & Eff. Mar. 2012$^h$\\
   &  &  &  &	&							   &$8.0\pm 0.9^f$  &$17.2\pm 2.0^f$  &0.020$^f$ & ATCA Sep. 2009\\
 $3_0-2_1A^+$    & 156.602413(10)$^{c}$  & $6 \times 10^{-8}$  &  0.02 & 83.042079    & -2.7  &$9.5\pm 1.5$& $11.1\pm 3.8$ &0.019 & IRAM Aug. 2012\\
 $1_{-1}-1_0E$ & 157.270851(10)$^{c}$  & $6 \times 10^{-8}$  &  0.02 & 83.396534    & -3.5  &  $10.5 \pm 0.7^g$  &   $15.4 \pm 0.4^g$  & 0.08 &IRAM Aug. 2012\\
 $2_{-1}-2_0E$ & 157.276058(10)$^{c}$  & $6 \times 10^{-8}$  &  0.02 &  83.399295   & -3.5  & '' & '' &   & IRAM Aug. 2012\\
 $3_{-1}-3_0E$ & 157.272369(10)$^{c}$  & $6 \times 10^{-8}$  & 0.02  & 83.397339   & -3.5  &  '' &      ''          &   & IRAM Aug. 2012\\
 $1_0-1_1A^{+/-}$    & 303.36689(5)$^{d}$  & $2 \times 10^{-7}$  &  0.06&160.86736      & -1.9  &$8.8\pm 1.0$ & $11.7\pm 2.4$  & 0.12 & IRAM Aug. 2012\\
 $3_0-4_1A^+$    & 492.278713(50)$^{e}$  & $1 \times 10^{-7}$  &  0.03 &261.042259    & -1.6  &$11.7\pm 0.3$  & $15.1 \pm 0.7$ & 0.04 & ALMA Jun. 2012  \\
\tableline
\end{tabular}
\\
$^a$Breckenridge and Kukolich (1995) \cite{Breckenridge1995}.
$^b$Heuvel and Dynamus (1973) \cite{Heuvel1973}.
$^c$M\"{u}ller\,\emph{et\,al.} \cite{Muller2004}.
$^d$Sastry\,\emph{et\,al.} \cite{Sastry1985}.
$^e$Herbst\,\emph{et\,al.} \cite{Herbst1984}.
$^f$Values from \cite{Ellingsen2012}, $V_{\rm HEL}$ is converted to $V_{\rm LSR}$ via $V_{\rm LSR} - V_{\rm HEL} = 12.432$ km s$^{-1}$.
$^g$Three components fitted jointly (see text).
$^h$Data used in \cite{Bagdonaite2013} where the values of the widths were stated as FWHM/$\sqrt(2ln2)$.
\end{center}
\end{table*}

\begin{figure}[ht!]
\includegraphics[scale=.3, trim=0cm 1.5cm 1cm 1.5cm]{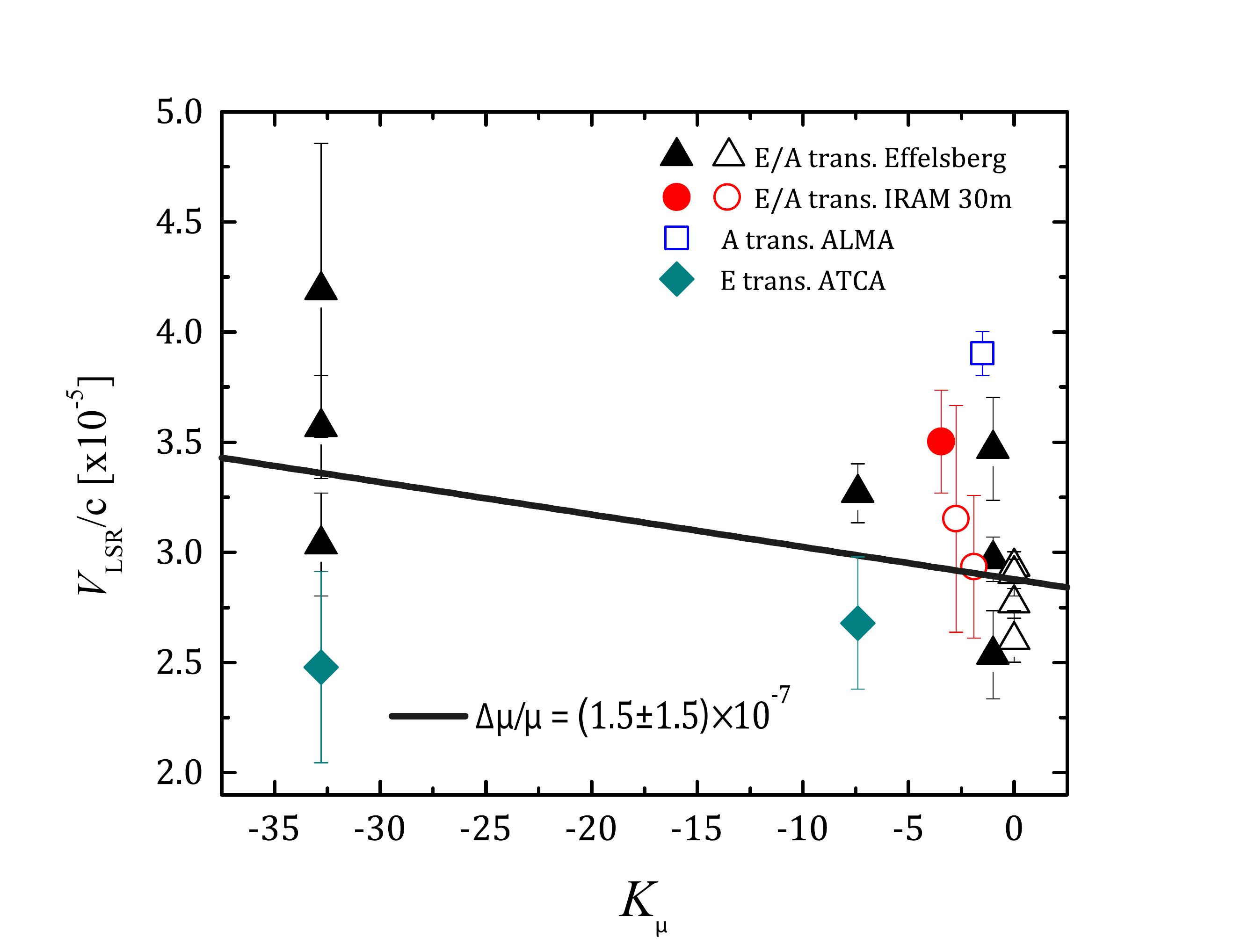}
\caption{\small{The positions of the ten observed methanol lines (represented by $V_{\rm LSR}$/c with respect to $z=0.88582$) are plotted versus their sensitivity coefficients, $K_{\mu}$. Note that some transitions were observed multiple times, hence, 17 data points are displayed. The slope of a straight line fitted to the dataset represents $-\Delta\mu/\mu$. The $A$ transitions observed at Effelsberg are offset from $K_{\mu} = -1$ to 0 for clarity.}}
\label{Fig4}
\end{figure}

\begin{figure}[ht!]
\includegraphics[scale=.4, trim=2cm 1cm 3cm 1cm ]{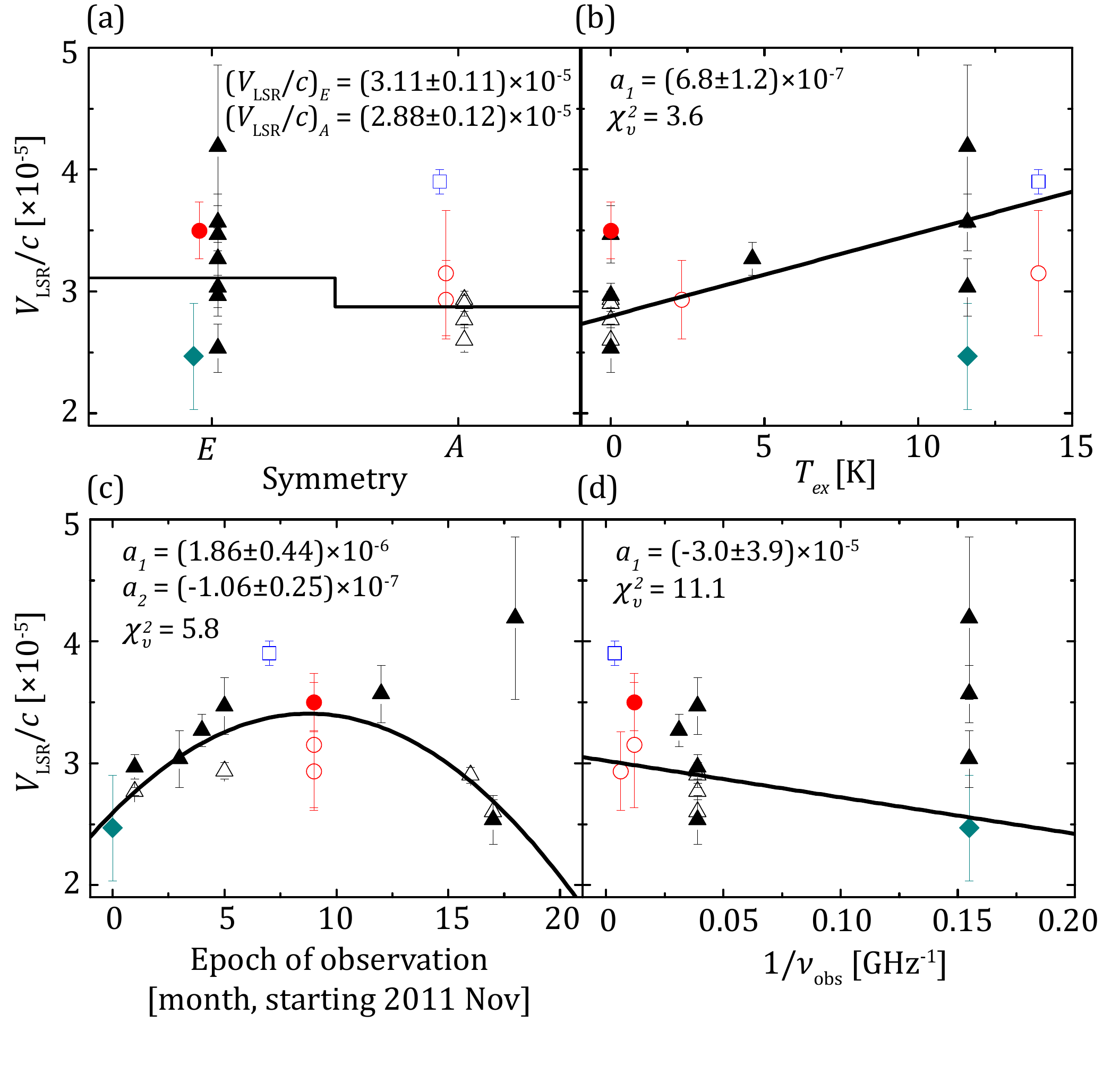}
\caption{\small{The positions of 16 observed methanol absorptions (leaving out the data point measured in 2009) as a function of various parameters: (a) symmetry, (b) excitation temperature of the lower level, (c) epoch of observation, (d) observed frequency. The color coding is the same as in Fig. \ref{Fig4}. The solid lines show a fit using a linear or quadratic function of form $y = a_{0} + a_{1}x + a_{2}x^2$.}}
\label{Fig5}
\end{figure}

Table~\ref{tbl-1} contains results from the fitted spectra of all methanol lines observed, as well as laboratory frequencies~\cite{Breckenridge1995,Heuvel1973,Muller2004,Sastry1985,Herbst1984} with their uncertainties, and the calculated $K_{\mu}$ coefficients~\cite{Jansen2011a}. The quoted optical depths $\tau$ are related to the measured absorption line intensities $I_{\nu}$ via $\tau = -\ln (1-I_{\nu}/f_{\rm c} I_{\rm bg}) $
where $I_{\rm bg}$ is the total continuum intensity and $f_{\rm c}$ is the continuum source covering factor. For unresolved single-dish observations, we adopt $f_{\rm c} = 0.4$ \cite{Henkel2009} for the SW sightline of PKS1830$-$211. 

The RADEX code allows us to derive a column density of methanol from the present set of measurements. Assuming a molecular hydrogen density of $2 \times 10^3$ cm$^{-3}$ and a kinetic temperature of 80 K (values discussed in~\cite{Henkel2009},\cite{Menten2008}, and \cite{Muller2013}), FWHM linewidths and optical depths as fitted to observations, and a $T_{\rm CMB} = 2.728(1 + z) = 5.145$ K for the temperature of the cosmic microwave background radiation at $z = 0.88582$, this yields a total column density of $~1.6 \times 10^{14}$ cm$^{-2}$ (adding $E$ and $A$ type methanol column densities).

All the detected methanol transitions have been included in the analysis. In addition, Table \ref{tbl-1} includes two data points resulting from observations with the Australia Telescope Compact Array (ATCA), presented in literature: the $2_{-1}-1_0E$ transition in \cite{Muller2011} observed in September 2009, and the $3_{-1}-2_0E$ transition in \cite{Ellingsen2012} observed in November 2011. We adopt line positions from the latter work.

The velocities between different transitions are interrelated via $V/c= -K_{\mu}\Delta\mu/\mu$, where $c$ is the speed of light (also see Eq.~\ref{eq1}). The error bars in the velocity positions reflect the statistical uncertainties from Gaussian profile fitting to the observed spectra. The uncertainty in the laboratory frequencies, expressed on a velocity scale in Table~\ref{tbl-1} with $\Delta v_{\rm D}$, is small ($<$14$\%$) compared to the uncertainty obtained from the astrophysical spectra and is treated as negligible. The uncertainties arising in the calibration of the frequency scale at the four observatories is of the order of 20 m s$^{-1}$ or less \cite{Levshakov2010, vandertak2009}. The result of the statistical analysis deducing $\Delta\mu/\mu$ is shown in Fig.~\ref{Fig4}. A fit to the 17 data points delivers a constraint of $\Delta\mu/\mu = (1.5\pm1.5)\times10^{-7}$ with a reduced $\chi^2_{\nu} = 10.2$, where the number of degrees of freedom, $\nu$, is 15. The large value of the goodness-of-fit measure $\chi^2_{\nu}$ indicates that the spread of the data is larger than expected from their errors. As a consequence, the addition of new data to our previous set \cite{Bagdonaite2013} did not lead to an expected improvement of the constraint. Assuming we do not underestimate the errors from the observations, the rather large value of $\chi^2_{\nu}$ suggests an underlying systematic effect that is not taken into account. In the following we address possible candidates.




\emph{Chemical segregation.}
Spatial segregation of molecular species within an absorbing galaxy may mimic or hide a variation of $\mu$. Since we rely on a single molecular species chemical segregation cannot be an issue, however, it may be possible that $E$ and $A$ type methanol are displaced spatially as suggested in \cite{Bagdonaite2013}. In panel (a) of Fig.~\ref{Fig5}, the data are grouped by their symmetry. The averages of the $A$ and $E$ transitions agree within their uncertainties. Furthermore, a two-dimensional linear regression (LR) analysis with $K_{\mu}$ and $E$/$A$ symmetry as independent variables results in an increase of $\chi_{\nu}^2$, if compared to a fit with $K_{\mu}$ alone, thereby ruling out a possible $E$/$A$ segregation. This new result of the extended study implies that all data pertaining to methanol can be included in a $\mu$ variation analysis.

\emph{Temperature dependence.}
The spread in the line positions might be ascribed to an inhomogeneous temperature distribution in the absorbing cloud. In panel (b) of Fig.~\ref{Fig5}, the $V_{\rm LSR}/c$ values are plotted as a function of the excitation energy of the lower level. The solid line shows a linear fit to the data, indicating a correlation between the measured line positions and the excitation energy. When we include the excitation energy as an independent variable in a two-dimensional LR analysis, we find a constraint of $\Delta\mu/\mu = (-1.3\pm1.0)\times10^{-7}$ with $\chi^2_{\nu} = 3.4$ where $\nu = 14$. The obtained limit of a time variation of $\Delta\mu/\mu$ changes only marginally when effects of the excitation energy are included, i.e., the $\Delta\mu/\mu$ is not strongly correlated to the excitation energy in our data set, however, the $\chi_{\nu}^2$ is significantly reduced.


\emph{Time variability of the background source.}
The strength of radio absorption lines towards the SW image was found to vary by a factor of $\sim2$ in a time span of three years, which was ascribed to the intensity changes in the background continuum source~\cite{Muller2008}. In particular, observations of an optically thick HCO$^+$ transition have shown an absorption profile, composed of several components, evolving with time. This phenomenon was explained by morphological changes in the background blazar. Thus, comparing line profiles from various time periods may lead to potential errors in line positions, especially if the velocity structure is underrepresented by the fitting \cite{Murphy2008}. 
In panel (c) of Fig. \ref{Fig5} we show $V_{\rm LSR}/c$ as a function of observation epoch. An indication of an oscillating behaviour can be found but as we observe (less than) one period of this oscillation we choose to fit a quadratic time dependence instead of a sinusoidal one. A two-dimensional LR analysis with the observation epoch and $K_{\mu}$ as independent variables results in $\Delta\mu/\mu = (0.7\pm1.1)\times 10^{-7}$ with $\chi_{\nu}^2 = 5.6$ with $\nu = 12$. Again, although our data show a dependence on observation epoch, this does not change the resulting limit on a time variation of $\Delta\mu/\mu$  if compared to the one from a $K_{\mu}$-only fit. In a three-dimensional analysis where both the time variability and the temperature dependence are taken into account, a constraint of $\Delta\mu/\mu = (-1.0\pm0.8)\times 10^{-7}$ is delivered with $\chi_{\nu}^2 = 2.4$, where $\nu = 11$.


\emph{Frequency dependence.} 
Previous studies have shown that the size of the south western image of the PKS1830$-$211 background blazar changes with frequency \cite{Guirado1999} and it exhibits a chromatic substructure, i.e. a different apparent position of the core at different frequencies, known as a core-shift effect. In particular, the angular separation $\Delta\theta$ of two sightlines at observed frequencies $\nu_1$ and $\nu_2$ is estimated to be $\Delta\theta = \Omega(1/\nu_1 -1/\nu_2)$ where $\Omega \sim 0.8$ mas GHz \cite{martividal2013}. No correlation is found between $V_{\rm LSR}/c$ and 1/$\nu$ (Fig. \ref{Fig5}(d)) but note that the sensitivity of the lines used in our study follow a 1/$\nu$ dependence \cite{Jansen2011a}, hence, $K_{\mu}$ and 1/$\nu$ are correlated and cannot be fitted simultaneously. According to \cite{martividal2013}, the core-shift effect in the PKS1830$-$211 system may introduce a shift of $\sim1$~km~s$^{-1}$ between lines at $\nu_{\rm obs}\sim 6$\,GHz and higher frequencies, which would translate into an uncertainty in $\Delta\mu/\mu$ of $1\times 10^{-7}$. As it is not possible to estimate the size of this effect based on the current data set, we adopt the latter value as a systematic uncertainty of $\Delta\mu/\mu$. In order to constrain the core-shift effect based on methanol only it would be desirable to add low frequency methanol absorption that have positive or small $K_{\mu}$. Attempts to observe the $3_{0}-4_{-1}E$ transition at $\nu_{\rm obs} \sim 19.2$\,GHz with $K_{\mu}=+9.7$ using the Effelsberg telescope were unsuccessful.

In conclusion, we present a test on a possible variation of $\mu$ at more than half of the age of the Universe using methanol absorption lines detected in a foreground galaxy towards the PKS1830$-$211 blazar. The methanol method, being the most sensitive probe for time-variations of $\mu$, can only be applied to this object as it is the only place in the far distant Universe where the methanol molecule has been detected so far. For this reason we focused all efforts on observing this object and used different telescopes to collect a total of 17 data points for ten different absorption lines from which a statistical constraint of $\Delta\mu/\mu = (1.5 \pm 1.5) \times 10^{-7}$ is derived. The large data set greatly enhances the understanding of previously unaddressed systematic effects and allows for a robust analysis. The suggested systematic on $E/A$ chemical segregation \cite{Bagdonaite2013} has now been discarded. The analysis reveals that effects pertaining to temperature inhomogeneity of the absorbing cloud and time variability of the background source result in a larger scatter than expected from the error of the individual transitions. 
By including the underlying systematic effects as independent variables in a multi-dimensional linear regression analysis, we obtain $\Delta\mu/\mu = (-1.0 \pm 0.8_{\rm stat} \pm 1.0_{\rm sys}) \times 10^{-7}$. Translated into a rate of change this corresponds to $\dot{\mu}/\mu < 2 \times 10^{-17}$ yr$^{-1}$, which is equally constraining as the bound on a varying constant obtained with the best optical clocks in laboratory experiments~\cite{Rosenband2008}.



\acknowledgments
This work is supported by the FOM-program ``Broken Mirrors \& Drifting Constants" and by FOM-projects 10PR2793 and 12PR2972. W.U. and P.J. acknowledge support from the Templeton Foundation, and H.L.B. acknowledges support from NWO via a VIDI grant. The research leading to these results has received funding from the European Commission Seventh Framework Programme (FP/2007-2013) under grant agreement No 283393 (RadioNet3). This paper makes use of the following ALMA data: ADS/JAO.ALMA$\#$2011.0.00405.S. ALMA is a partnership of ESO (representing its member states), NSF (USA) and NINS (Japan), together with NRC (Canada) and NSC and ASIAA (Taiwan), in cooperation with the Republic of Chile. The Joint ALMA Observatory is operated by ESO, AUI/NRAO and NAOJ. Based on observations carried out with the IRAM 30-m telescope. IRAM is supported by INSU/CNRS (France), MPG (Germany) and IGN (Spain). Staff of the Effelsberg, IRAM 30-m and ALMA radio telescopes are thanked for their support.

\end{document}